\documentclass[twocolumn,showpacs,showkeys,amsmath,amssymb]{revtex4}

\usepackage{graphicx}% Include figure files
\usepackage{dcolumn}% Align table columns on decimal point
\usepackage{bm}% bold math

\begin{document}

\preprint{LEI 05--06}

\title{Analysis of laboratory nucleosynthesis products}

\author{S.~V.~Adamenko}

\author{A.~S.~Adamenko}

\affiliation{Electrodynamics Laboratory `Proton--21', Kiev, Ukraine}

\email{enr30@enran.com.ua}

\begin{abstract}%
We present the results of the experimental study on synthesis of a
wide range of isotopes in a superdense plasma. The initial
conditions necessary for plasma bunch formation were provided by
specially organized coherent impact on a solid target with a total
energy up to 1\,kJ. More than 4000 shots were performed with
various targets made of light, medium, and heavy elements.
Subsequent analysis of the products of the target explosion
reveals the presence of a wide range of elements absent in the
initial materials. Elements with nuclei three and more times
heavier than the nucleus of the target main element are detected
in the products. The isotopic composition of the produced elements
significantly differs from the natural one. The presence of
unknown superheavy elements at the border of the periodic table
and beyond it was detected by several different spectroscopic
methods of elemental and isotopic analyzes.%
\end{abstract}

\copyright{Electrodynamics Laboratory `Proton--21'}

\maketitle

\section*{Introduction}
This work summarizes the results of analysis of the nuclear
transmutation pro\-ducts in the supercompressed substance obtained
in 1999--2003 by the Electrodynamics Laboratory `Proton--21'
(Ukraine, Kiev) that was established to develop a novel approach
for radioactive waste utilization.

The project was based on the idea that it is possible to create a
special kind of disturbance impact on a solid substance that will
induce, under certain conditions, the self--organizing avalanche
process of supercompression up to the collapse state at which the
conditions for collective multiparticle transmutation reactions of
radioactive elements into stable ones will become possible, owing
to the redistribution of neutrons in the substance volume and
increase of their concentration at the center of the collapse. A
few experimental setups have been built in the laboratory. The
first successful experiment was performed on February 24, 2000. By
March, 2003, 4037 dynamic compressions of solid targets were
performed leading to specific explosions and the radial dispersion
of a transformed target material from the collapse zone.

\section{Multinuclear reactions
in a superdense substance}
\subsection{Experiment}
The main experimental setup realizes the consecutive space--time
energy compression and is capable to impact on the solid target up
to 1\,kJ of energy during the pulse time less than 100\,ns. The
electron beam is used as an initial carrier of the concentrated
energy. On the final stage, the  substance of target is compressed
up to a density of $>10^{26}$\,cm$^{-3}$, while the estimated
power density in the collapse volume exceeds $10^{22}
$\,W\,cm$^{-3}$.

The target explodes as a result of dynamic compression. At
explosion an ultrabright quasipoint X--ray source is formed with
registered energy range from $2 \dots 3$\,keV up to 10\,MeV and
more, with maximum intensity in the range $10\dots30$\,keV.
Products of a target explosion precipitate on the internal walls
of concentrator and on the special accumulating screen
--- metallic disk 1\,cm in diameter. The precipitated products
have the form of irregularly scattered drops, beads, films, etc.

\subsection{Analysis}

Precipitated products were analyzed to determine their isotopic
and elemental compositions. The analytical methods employed were:
\begin{itemize}
\item electron probe microanalysis (EPMA);%
\item Auger--electron spectroscopy (AES);%
\item secondary ion mass--spectrometry (SIMS);%
\item laser mass--spectrometry (LMS);%
\item Rutherford backscattering of accelerated alpha--particles
(RBS).
\end{itemize}

As a rule, after the experiment, elemental analyzes of the surface
 and surface layer up to 3\,$\mu$m deep of
each accumulating screen (made of chemically pure material:
Cu~(99.99\,\% pure), Ag~(99.99\,\%), Ta~(99.97\,\%),
Pb~(99.75\,\%), etc.) were performed. Minimal number of
microprobes was from 10 to 20, some regions (spots) were also
analyzed by several methods.

By March 12, 2003, altogether 14520 analyzes were performed and
registered. The summary of the analytical methods and number of
analyzes are listed in Table~\ref{tab:anal}.

\begin{table}[h]%
\centering%
\caption{List of analytical methods and number of
analyzes.}%
\vspace*{10pt}
\begin{tabular}{l|r|r}

\hline
{} &{} &{} \\[-2.5ex]
Method              & Number of & Number of \\
                    &   samples &    probes \\
\hline
\hline
EPMA                &   541 & 9231  \\
LMS                 &   20  &  297  \\
AES                 &   25  &  474  \\
SIMS                &   24  &  399  \\
RBS                 &   40  &   40  \\
EPMA+LMS            &   38  & 1227  \\
EPMA+AES            &   44  & 1522  \\
EPMA+SIMS           &   21  &  619  \\
EPMA+LMS+AES        &    4  &  164  \\
EPMA+LMS+SIMS       &    2  &   57  \\
EPMA+AES+SIMS       &    7  &  316  \\
EPMA+LMS+AES+SIMS   &    1  &   43  \\
LMS+AES             &    1  &   29  \\
AES+SIMS            &    2  &  102  \\
\hline
{} &{} &{} \\[-2.5ex]
total EPMA          &   658 &11578  \\
total LMS           &   66  & 1001  \\
total AES           &   84  & 1359  \\
total SIMS          &   57  &  542  \\
total RBS           &   40  &   40  \\
\hline
{} &{} &{} \\[-2.5ex]
TOTAL               &   770 &14520  \\
\end{tabular}\label{tab:anal}
\vspace*{-13pt}
\end{table}

\subsection{Main results of elemental composition
analyzes after one of the typical experiments}

Analyzes show that the products of impact explosion precipitated
on the accumulating screen contain a wide spectrum of the light,
medium and heavy chemical elements that are absent in comparative
quantities or concentrations in the initial materials involved in
the process of nuclear transformation.

This fact is illustrated in Table~\ref{adtab1}, where the
quantities of impurities in chemically pure copper (used for the
target and accumulating screen)  are compared with those detected
in the products precipitated on the accumulating screen after one
of the typical experiments. Correlation ratios between the
concentration of impurities in the target material and their
quantities in the precipitated products have near zero or negative
values.

The composition of the initial environment and target materials
were measured with a mass--spectrometer of glow discharge VG~9000
(Thermo Elemental), with sensitivity from 100\,\% to ppt level in
a single analysis.

Three different instruments were used for analysis of the
elemental composition of the accumulating screen surface layer
after the
target impact explosion:%
\begin{itemize}
\item Auger spectroscope JAMP--10S (JEOL, Japan) --- 21 spots
      $<$1.0\,$\mu$m in diameter and 0.002\,$\mu$m deep were analyzed;%
\item EPMA analyzer REMMA 102 (Ukraine) --- 113 areas
      11$\times$11\,$\mu$m and 3.0\,$\mu$m deep, and 417 extrinsic
      beads of different shape on the surface of the accumulating
      screen.%
\item SIMS analyzer IMS\,4f (CAMECA, France) --- 5 areas
      250$\times$250\,$\mu$m and 0.4\,$\mu$m deep (because SIMS is
      the destructive method, it was used after Auger and EPMA);%

\end{itemize}

As can be seen from Table~\ref{adtab1}, both the light elements
(with the mass number less than the initial copper) and the heavy
ones are present. It should be emphasized that the elements with
atomic masses exceeding two masses of the initial element
(elements with mass number $Z>58$) were detected as well.
Quantities of the mentioned elements by a few orders of magnitude
exceed impurities in the initial material.

\begin{center}
\begin{table}[h]
\centering \caption{Number of atoms in the surface layer of
accumulating screen.} \vspace*{10pt}
\begin{tabular}{c|r|r|r}
\hline
{} &{} &{} &{}\\[-2.5ex]
Element & $Z$ & Init Cu target & Accum. screen\\
\hline
\hline
Li&   3&       1.7\,E+12&     6.0\,E+11\\%
Be&   4&       6.1\,E+11&     1.3\,E+14\\%
B&    5&       2.1\,E+12&     4.1\,E+13\\%
C&    6&          ---   &     9.5\,E+17\\%
N&    7&          ---   &     1.1\,E+15\\%
O&    8&          ---   &     4.3\,E+15\\%
Na&   11&      6.5\,E+13&     1.3\,E+16\\%
Mg&   12&      3.6\,E+13&     3.3\,E+15\\%
Al&   13&      3.9\,E+14&     3.3\,E+17\\%
Si&   14&      3.8\,E+13&     9.8\,E+16\\%
P&    15&      6.5\,E+14&     2.0\,E+16\\%
S&    16&      3.4\,E+14&     1.2\,E+17\\%
Cl&   17&      2.4\,E+10&     1.5\,E+17\\%
K&    19&         --- &       5.3\,E+16\\%
Ca&   20&      3.2\,E+14&     1.8\,E+16\\%
Ti&   22&      2.3\,E+12&     3.8\,E+15\\%
V&    23&      1.1\,E+11&     9.1\,E+13\\%
Cr&   24&      3.3\,E+12&     2.5\,E+15\\%
Mn&   25&      2.4\,E+13&     1.5\,E+15\\%
Fe&   26&      1.3\,E+15&     8.7\,E+16\\%
Co&   27&      1.0\,E+12&     3.9\,E+14\\%
Ni&   28&      3.8\,E+14&     2.0\,E+14\\%
Zn&   30&      5.5\,E+13&     7.5\,E+16\\%
Y&    39&      1.9\,E+10&     2.0\,E+14\\%
Zr&   40&      5.9\,E+10&     2.8\,E+13\\%
Ag&   47&      8.5\,E+13&     6.4\,E+15\\%
Cd&   48&      1.1\,E+12&     2.2\,E+15\\%
In&   49&      9.7\,E+11&     1.9\,E+15\\%
Sn&   50&      2.0\,E+13&     1.6\,E+16\\%
Te&   52&      8.6\,E+12&     1.4\,E+15\\%
Ba&   56&      3.2\,E+11&     2.4\,E+15\\%
La&   57&      1.4\,E+10&     7.2\,E+14\\%
Ce&   58&      2.2\,E+10&     2.5\,E+15\\%
Pr&   59&      2.6\,E+10&     1.5\,E+14\\%
Ta&   73&         ---   &     4.2\,E+15\\%
W&    74&      3.1\,E+11&     2.3\,E+16\\%
Au&   79&      1.0\,E+11&     5.8\,E+15\\%
Pb&   82&      2.5\,E+13&     2.0\,E+17\\%
\hline
\multicolumn{2}{c|}{TOTAL}&  3.7\,E+15&2.2\,E+18\\
\end{tabular}\label{adtab1}
\vspace*{-13pt}
\end{table}
\end{center}

To get quantitative estimations for the number and the
concentration of deposited atoms, the statistical analysis was
performed for the following layers:
\begin{enumerate}
\item Thin surface layer about 0.002\,$\mu$m;%
\item Intermediate layer from 0.002\,$\mu$m to 0.4\,$\mu$m;%
\item Rest of the material --- layer from 0.4\,$\mu$m to
3.0\,$\mu$m (maximum depth of analyzes).
\end{enumerate}

As the result of statistical evaluation, the concentrations in the
above--mentioned layers were obtained for each element. The
validation criterion was the accordance between the calculated
element concentration and concentration measured by all applied
instruments.

Statistical analysis of the quantity of deposited atoms revealed the following:%
\begin{itemize}
\item Deposited atoms ($2.8\times10^{16}$\,atoms) concentration is
      maximum in the surface layer 0.002\,$\mu$m thick (1st
      layer) and the concentration of copper (main target and screen material)
      is lower than 10\,\%.

\item Main quantity of deposited elements
      ($1.3\times10^{18}$~atoms) and the majority of the elements
      with masses exceeding two masses of the initial material
      (i.e. Cu) are present in the 2nd layer (From 0.002 to 0.25\,$\mu$m).

\item Quantity of deposited atoms in the 3rd layer
      (0.25\,$\mu$m and down to the limit of the analyzer, 3.0\,$\mu$m)
      is $3.1\times10^{17}$~atoms.

      Total number of deposited atoms, excluding copper, present
      in the substance up to 3.0\,$\mu$m deep was calculated as sum of the
      atom quantities in the volumes of each layer and equals
      $1.6\times10^{18}$~atoms. Change of the thickness of the 2nd layer
      containing the main part of deposited elements from 0.2\,$\mu$m to
      0.4\,$\mu$m gives the range $(1.4\dots 2.4)\times10^{18}$
      for the total number of deposited atoms.

\item Quantity of deposited atoms in randomly dispersed
      particles on the surface of the accumulating screen is equal to
      $6.0\times10^{17}$\,atoms.
\end{itemize}

The total number of deposited atoms, excluding copper, after the
experiment in comparison to the initial concentration is shown in
Table\,\ref{adtab1}.

The same estimation for the synthesized atoms was derived from the
`marked target' experiments. The targets for these experiments
were fabricated from radioactive cobalt ($^{60}$Co). After an
impact, the system activity (number of radioactive decays per
second) decreased by the amount equal to the transmutation of
$10^{18}$ atoms of the active target volume --- intensity of the
$^{60}$Co spectral lines decreased while no other radioactive
isotopes line did appear.

\subsection{Isotopic composition
analyzes\\ of accumulating screens}

To confirm that we deal with the nucleosynthesis process, the
isotopic composition of the products on the surface of the
accumulating screens was analyzed. The stimulus for this was the
well--known fact that the isotopic composition of any element is
almost identical throughout the Solar system (see, e.g.
\cite{iupac,wasser}).

\begin{figure}[h]
\centering
\includegraphics[width=7.5cm,keepaspectratio]{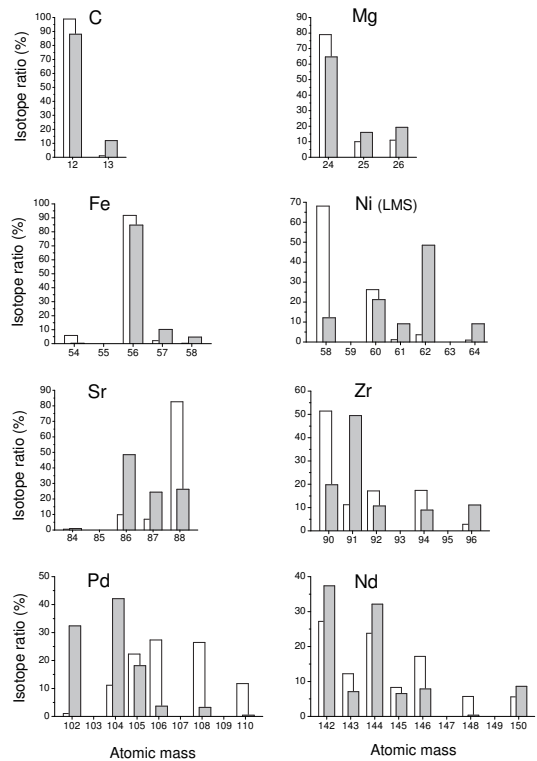}

\caption{Isotopic composition of some elements measured with LMS
(indicated) and SIMS (others). Natural composition is depicted
with empty bars, the composition of synthesized elements with
hatched bars. \label{fig_ad1}}
\end{figure}

Analyzes of the isotopic composition were done with two different
methods, i.e., LMS and SIMS (see Table~\ref{tab:anal}). As was
found out, most of the analyzed spots of an accumulating screen
had an isotopic composition that differs from a natural one.
Examples of the isotopic composition for some elements are
presented in Fig.\,\ref{fig_ad1}.

As can be seen from this figure, the isotopic composition of the
elements synthesized from the copper target, differs significantly
from the natural one. Hence, this proves an artificial origin of
the detected nuclei.

\section{Synthesis of the elements at the border of
the periodic table and beyond it}

The presence of heavy elements (tantalum, tungsten, gold, and
lead) among the products of nucleosynthesis from the copper target
encouraged us to carry out the experiments with heavier targets.
In fact, since  lead ($A(\mathrm{Pb})\approx 207$) was produced
from the copper target ($A(\mathrm{Cu})\approx 64$), one should
expect in similar way  to get elements at the border of the
periodic table and even beyond it if a target made from heavier
elements is used.

For this aim, the experiments with the platinum, lead, and bismuth
targets were carried out. The results of analyzes revealed the
presence of long-living isotopes of chemical elements at the
border and even beyond the known part of periodic table.

To identify the elemental composition of the accumulating screen
surface, the Auger spectroscopy (AES) method was used. The method
is non-destructive and highly localized ($50\dots 100$\,nm). The
depth of analyzes is low ($1\dots 2$\,nm). The method ensures a
wide range of detectable elements (all but H and He) and a
relatively high sensitivity ($0.1\dots 1$\,at.\%).

The spectra obtained with the help of a Auger--microprobe
JAMP--10S (JEOL, Japan) typically contained Auger peaks of a
considerable amount of chemical elements that hindered their
identification. To identify the Auger peaks, the measurement was
performed in a wide range of energies ($30\dots3000$\,eV, energy
resolution $0.5\dots 1.2$\,\%) to cover the maximum number of
Auger peaks' series, and prolonged exposures (up to 3 hours) were
used to identify the low intensity peaks. The spectra were
recorded in differential form at the accelerating voltage of the
electron probe of 10\,kV and beam current
$10^{-6}\dots10^{-8}$\,A. Residual pressure in the sample chamber
was $5\times10^{-7}$\,Pa. Artifacts like electrical charging,
characteristic energy losses, and chemical shift were taken into
consideration. A standard concentration calculation program
supplied by the manufacturer (JEOL) was used for quantitative
analysis.

In the process of analysis, the unidentified peaks with energies
of 172, 527, 1096\,eV, and a doublet of 130 and 115\,eV were
registered on the surface of accumulating screens. These peaks do
not correspond to any of the catalogued peaks of chemical elements
and cannot be referred to any of the known artifacts. One of the
possible explanation of the mentioned peaks is the presence of
long-living transuranium elements.

\begin{figure}[h]
\centering
\includegraphics[width=7.5cm,keepaspectratio]{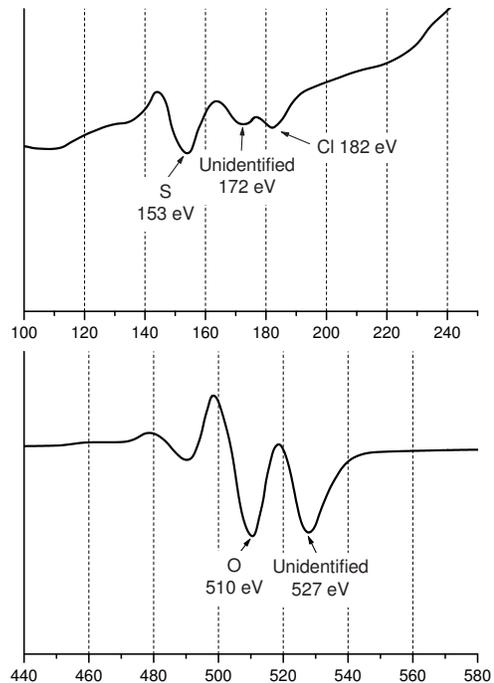}

\caption{Sections of Auger spectra with unidentified peaks.
\label{fig_ad2}}
\end{figure}

Two sections of the Auger spectrum (energies 172, 527\,eV) out of
six unidentified peaks are presented in Fig.\,\ref{fig_ad2}.

To provide elemental analysis of the accumulating screens surface
in a wide range of masses, a SIMS analyzer IMS\,4f (CAMECA,
France) was used.

Here are some key characteristics of the instrument:
\begin{itemize}
\item mass detection range --- up to 480;%
\item mass resolution ($M/\Delta M$) --- 2500;%
\item locality --- 0.2\,$\mu$m;%
\item primary ions: Cs$^{+}$, O$^{+}_{2}$, O$^{-}$, Ar$^{+}$;%
\item secondary beam stability (20 min): $\Delta I/I = 0.7\,\%$,
      \\$\Delta M/M = 7\times10^{-6}$.
\end{itemize}

A complex approach was used for elemental analyzes of the
accumulating screens surface and included studies of the
topography of chemical elements on the surface in the cluster
suppression mode (Offset) and high mass resolution mode
simultaneously with the photo film recording of the mass
distribution.

In the Offset mode, an additional voltage is applied and, owing to
the potential barrier, allows us to identify monoatomic ions even
in the low--mass resolution mode and, hence, to exclude clusters
and multicharged ions.

From the whole mass--spectrum, we consider as unidentified only
those peaks which (a) were not observed in the reference material
spectrum, (b) did not coincide with the other elements'
distribution on the screen surface, (c) offset voltage increase
led to an increase of the intensity in comparison with clusters'
peaks.

\begin{figure}[h]
\centering
\includegraphics[width=7.5cm,keepaspectratio]{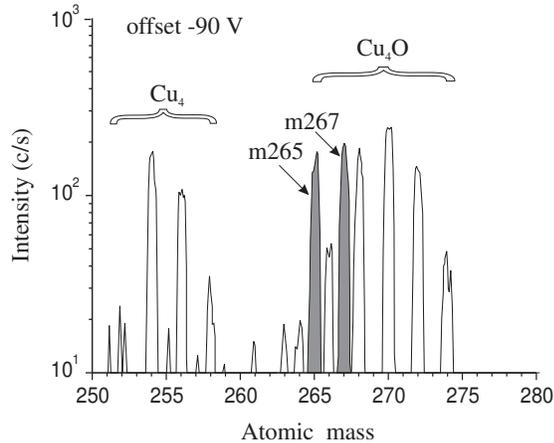}

\caption{Mass spectrum revealing the presence of unidentified
masses. \label{fig_ad3}}
\end{figure}

Typical mass spectrum in the mass range $250\dots 280$ obtained by
IMS\,4f is presented in Fig.\,\ref{fig_ad3}. Filled peaks,
corresponding to 265 and 267, are unidentified; 268, 270, 272, 274
are Cu$_{4}$O. Increasing the offset voltage led to a significant
decrease of the intensity for copper clusters and the negligible
fall of the intensity of 265 and 267 peaks indicating that they
belong to monoatomic ions. To exclude possible chemical elements
combinations, the photos of all present elements distributions
were done, and they show that none of the unidentified masses
coincide with any of them.

Up to date, 20 samples were thoroughly checked for unidentified
masses. On 17 samples, more than 100 unidentified masses were
found in the range $221\dots 475$. Most frequently were detected
the peaks of 271, 272, 330, 341, 343 masses. Accumulating screens
that were used repeatedly in experiments with copper targets also
revealed the presence of long-living isotopes in the mass range
$300\dots481$.

\begin{figure}[h]
\centering
\includegraphics[width=7.5cm,keepaspectratio]{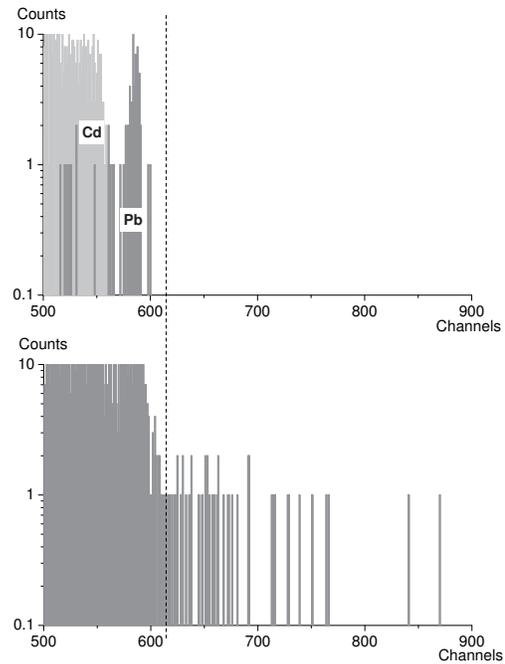}

\caption{Rutherford backscattering data; initial material (above)
and after the impact (below). \label{fig_ad4}}
\end{figure}

In addition to above--mentioned spectrometric methods, the samples
were examined for the presence of superheavy elements with the
most direct method~--- Rutherford backscattering.
The samples were irradiated with the beam of 27.2\,%
MeV alpha--particles accelerated in a U--120 cyclotron
\cite{Shvedov}. The energy spectrum of scattered projectiles
exhibited the scattering centers corresponding to mass numbers in
the range $200\dots 400$ (Fig.\,\ref{fig_ad4}).

Thus, different methods of analysis detected the presence of
unidentified superheavy elements among the products of artificial
nucleosynthesis.

\section{Summary of results and discussion}
We have presented the survey of experimental studies of
nucleosynthesis products obtained in experiments with the
supercompressed substance. The essence of results is as follows:
\begin{itemize}
\item In substance compressed up to a superhigh density
(presumably, $10^{26}$\,cm$^{-3}$), the process of nucleosynthesis
and transmutation occurs. This process takes place over a
macrovolume of the target substance
($2\dots5\times10^{18}$\,atoms/kJ).

\item Nuclear processes in a target are collective and
multiparticle that is proved by significant amount ($>10^{13}$) of
synthesized nuclei with masses more than two masses of the initial
nucleus.

\item No $\alpha$--, $\beta$--, $\gamma$--active isotopes were
observed in the products of laboratory nucleosynthesis, the
radiation intensity never exceeded the background intensity.

\item The activity of the targets marked with radioactive isotopes
was reduced after compression impact by the value equal to the
transmutation of $\sim10^{18}$\,atoms of the active target zone
(collapse zone) for every 1\,kJ of the driver energy.

\item Developed installation has high reproducibility in reaching
the conditions in a compressed substance necessary for the
ignition of collective multiparticle fusion--fission reactions of
the full spectrum of chemical elements.

\item The products of nuclear transmutation reveal the presence of
long--living isotopes of superheavy elements.

\end{itemize}

We consider that these reactions are, in some sense, similar to
the pycnonuclear reactions that take place under the specific
conditions of compact astrophysical objects (white dwarfs,
supernovas, outer shells of neutron stars, etc.).

The new challenge is the development of a method and a technique
for simultaneous measurement of both the atomic and nuclear
properties of superheavy atoms, i.e., nucleus charge and mass.

The results of our experiments allow us to suppose that the
synthesized nuclei are the product of clusterization in the
decaying superdense electron--nucleon plasma that corresponds to
maximization of the binding energy per nucleon, dependent  on the
substance density, on the one hand, and  on the neutron
concentration in the certain volume of clusterization, on the
other hand.

\section*{Acknowledgments} It is a pleasure for the authors to
thank \\ E.V.~Bulyak, I.A.~Kossko, Y.V.~Syten'ko,\\
S.A.~Shestakov, D.G.~Birukov, V.V.~Kovylyaev,\\
M.I.~Gorodyskiy, S.S.~Ponomarev, \framebox{G.A.~Zykov}

\end{document}